\documentclass{emulateapj}

  \usepackage{natbib}
  \bibpunct{(}{)}{;}{a}{}{,} 

\newcommand{\funits}{10$^{-16}$ erg~s$^{-1}$~cm$^{-2}$~\AA$^{-1}$}

\shorttitle{The night-sky at the Calar Alto Observatory.}
\shortauthors{S\'anchez et al.}

\begin{document}

\title{The night-sky at the Calar Alto Observatory}

\author{S.F~S\'anchez\altaffilmark{1}, 
J.~Aceituno\altaffilmark{1},
U.~Thiele\altaffilmark{1},
D.~P\'erez-Ram\'\i rez\altaffilmark{2,3},
J.~Alves\altaffilmark{1}
}

\altaffiltext{1}{Centro Astron\'omico Hispano Alem\'an, Calar Alto, (CSIC-MPG),
  C/Jes\'us Durb\'an Rem\'on 2-2, E-04004 Almeria, Spain }
\altaffiltext{2}{Departamento de F\'\i sica Aplicada, Facultad de Ciencias,
  Universidad de Granada,8 Fuentenueva s/n, 18071, Granada (Spain)}
\altaffiltext{3}{Centro Andaluz de Medio Ambiente,
Universidad de Granada. Junta de Andaluc\'\i a, Avda. del Mediterraneo s/n.
18006 Granada (Spain).}

\email{sanchez@caha.es}

\begin{abstract}
   
  We present a characterization of the main properties of the
  night-sky at the Calar Alto observatory for the time period between
  2004 and 2007. We use optical spectrophotometric data, photometric
  calibrated images taken in moonless observing periods, together with
  the observing conditions regularly monitored at the observatory,
  such as atmospheric extinction and seeing. We derive, for the first
  time, the typical moonless night-sky optical spectrum for the
  observatory. The spectrum shows a strong contamination by different
  pollution lines, in particular from Mercury lines, which
  contribution to the sky-brightness in the different bands is of the
  order of $\sim$0.09 mag, $\sim$0.16 mag and $\sim$0.10 mag in $B$,
  $V$ and $R$ respectively. Regarding the strength of the Sodium
  pollution line in comparison with the airglow emission, the
  observatory does not fulfill the IAU recomendations of a dark site.
  The zenith-corrected values of the moonless night-sky surface
  brightness are 22.39, 22.86, 22.01, 21.36 and 19.25 mag
  arcsec$^{-2}$ in $U$, $B$, $V$, $R$ and $I$, which indicates that
  Calar Alto is a particularly dark site for optical observations up
  to the $I$-band. The fraction of astronomical useful nights at the
  observatory is $\sim$70\%, with a $\sim$30\% of photometric
  nights. The typical extinction at the observatory is
  $\kappa_V\sim$0.15 mag in the Winter season, with little
  dispersion. In summer the extinction has a wider range of values,
  although it does not reach the extreme peaks observed at other
  sites. The analysis of the Winter and summer extinction curves
  indicates that the Rayleigh scattering is almost constant along the
  year. The rise of the extinction in the summer season is due to an
  enhance of the Aerosol extinction, most probably associated with an
  increase of dust in the atmosphere. The median seeing for the last
  two years (2005-6) was $\sim$0.90$\arcsec$, being smaller in the
  Summer ($\sim$0.87$\arcsec$) than in the Winter
  ($\sim$0.96$\arcsec$).

  We conclude in general that after 26 years of operations Calar Alto
  is still a good astronomical site. Its main properties are similar
  in many aspects to those of other major observatories where 10m-like
  telescopes are under operation or construction, being a natural
  candidate for future large aperture optical telescopes.

\end{abstract}



\section{Introduction}

The night sky brightness, the number of clear nights, the seeing,
transparency and photometric stability are some of the most important
parameters that qualify a site for front-line ground-based
astronomy. There is limited control over all these parameters, and
only in the case of the sky brightness is it possible to keep it at
its natural level by preventing light pollution from the immediate
vicinity of the observatory. Previous to the installation of any
observatory, extensive tests of these parameters are carried out in
order to find the best locations, maximizing then the efficiency of
these expensive infraestructures.  However, most of these parameters
are not constant along the time. An example of this can be seen in the
seeing evolution of the Paranal observatory, which is worse now than
when the decision to built it at its current location was
taken\footnote{http://www.eso.org/gen-fac/pubs/astclim/paranal/seeing/singstory.html}. This
is not an untypical situation.

We have started a program to determine the actual values of the main
observing conditions for the Calar Alto observatory.  The Calar Alto
observatory is located at 2168m height above the sealevel, in the
Sierra de los Filabres (Almeria-Spain) at $\sim$45 km from the
Mediterranean sea. It is the second largest european astronomical site
in the north hemisphere, just behind the Observatorio del Roque de los
Muchachos (La Palma), and the most important in the continental
europe. Currently there are six telescopes located in the complex,
three of them operated by the Centro Astronomico Hispano Aleman
A.I.E. (CSIC-MPG), including the 3.5m, the largest telescope in the
continental europe.

Along its 26 years of operations there has been different attempts to
characterize some of the main properties described before: (i) Leinert
et al. (1995) determined the sky brightness corresponding to the year
1990; (ii) Hopp \& Fernandez (2002) studied the extinction curve
corresponding to the years 1986-2000; (iii) Ziad et al. (2005)
estimated the median seeing in the observatory from a single campain
in may 2002. However, there is a need for a consistent study of all
these properties, spanning over a similar time period.

In this article we study the main characteristics of the night-sky at
the observatory including: (i) the night-sky spectrum, identifing the
natural and light pollution emission lines and their strength, (ii)
the moonless night-sky brightness in different bands, (iv) the
extinction and its yearly evolution and (v) the atmospheric seeing and
its yearly evolution. The study is limited to the last four years,
with mostly corresponds to a period of minimun solar
activity\footnote{http://www.ngdc.noaa.gov/stp/SOLAR/ftpsolarradio.html}
(which strongly affect several sky properties, like night-sky
brightness and airglow). The derived main properties have been
compared with similar properties at other observatories.

The structure of this article is as follows: in Section \ref{data} we
describe the dataset collected for the current study, including a
description of data and the data reduction; in Section \ref{ana} we
show the analysis performed over the different types of data and the
results derived for each one; in Section \ref{conc} we summarize
the main results and present the conclusions.

\section{Description of the Data}
\label{data}

In order to understand the properties of the night-sky emission at the
Calar Alto Observatory we collected different observational data,
including both imaging and spectroscopic data.  Since none of the data
were obtained directly for this study, we scanned thoroughly the
arquived data to acquire a data set with a sufficient degree of
homogeneity.

\subsection{Spectroscopic data}
\label{spec_data}

Spectroscopic data were obtained to determine the mean properties of
the night-sky spectrum of moonless nights at the observatory. In Calar
Alto spectrographs are normally mounted in bright and gray nights,
while dark nights are more frequently allocated for deep imaging
programs. Thus, it is somehow difficult to find spectrocopic data
taken during dark nights. The most frequently mounted spectrographs at
the Calar Alto observatory are CAFOS at the 2.2m ($\sim$50\% of the
allocated time) and PMAS \citep{roth05} at the 3.5m telescope
($\sim$30\% of the allocated time). PMAS is an integral field unit
with two different setups, a lensarray with a reduced field-of-view
(16$\arcsec$$\times$16$\arcsec$ in its largest configuration), and a
wide fiber-bundle that covers a field-of-view of
72$\arcsec$$\times$64$\arcsec$ with 331 individual fibers of
2.7$\arcsec$ diameter each one (PPAK, Kelz et al. 2006). This latter
configuration is particularly interesting to study the properties of
the sky emission, since in a single shot centred on a calibration star
(or a science target) a substantial fraction of the field-of-view
samples the sky. Its large aperture, and the possibility of performing
self-calibration, allows one to obtain spectrophotometric calibrated
and high signal-to-noise spectra of the night-sky emission even with
reduced exposure times. Evenmore, this instrument is frequently
mounted with a low-resolution grating (V300), which covers a
wavelength range of $\sim$3500\AA\ with a spectral resolution of
$\sim$10\AA\ (FWHM). This is also very convenient to obtain spectra of
the sky emission in all the optical wavelength range. Restricting
ourselves to the same instrument and configuration ensures the
homogeneity of the data.

There were 14 clear moonless nights when the instrument was mounted using this
configuration in the period between January 2005 and December 2006. A sample
of 23 observations, including night-sky emission spectra, were selected from
the data taken that nights. The sample was selected by including only
observations of calibration stars and/or small-size and faint targets (eg.,
High-z Ly-$\alpha$ emitters, S\'anchez et al. 2007c), with most of the PPAK
field-of-view sampling sky emission. In addition, only observations near the
zenith, with an airmass lower than 1.5, were included in the sample. In all
the cases the observations were taken far away the ecliptic and the galactic
plane. The data were then reduced using R3D (S\'anchez et al. 2006), following
the standard steps for fiber-based IFS data. Once reduced, the sky spectra
were extracted from the frames using E3D \citep{sanc04}, by selecting areas
clean of objects within the field-of-view. Although the signal-to-noise level
of each individual sky spectrum is somehow different, depending on the
exposure time and the number of selected fibers to extract the spectra, in all
the cases is good enough for the purposes of this study.

Not all the observations during a moonless night are equally dark. The
darkness of an observation depends strongly on the time distance of the
corresponding night from the full moon, the presence of cirrus, dust and local
contamination when pointing towards highly populated areas (like Almeria,
towards the south of the observatory), the zenith distance, and even more the
time distance from the twilight. Thus, from the original dataset we did not
consider those spectra which intensity at 5200\AA\ was larger than 0.3 \funits
(V$\sim$20.2 mags). We were then left with a final sample of 10 individual
spectra obtained in 7 different nights, representative of the typical dark
observation in a moonless dark night at Calar Alto. This sample still
comprises a wide range of brightness, and only one (02/06/2005) can be
considered completely dark, corresponding to a new moon night. Table
\ref{tab_data} shows the final list of nights, together with the wavelength
range covered by the spectra each night.

\subsection{Imaging data}
\label{img_data}

Multiband imaging data are collected to characterize the
sky-brightness of moonless nights at the observatory. The search was
restricted to CAFOS covering a similar period of time of the selected
spectroscopic dataset, to preserve the homogeneity of the data. Only
exposures on calibration fields were selected to perform a self
flux-calibration, avoiding the possible errors due to changes in the
atmospheric condition between calibration and measurement. Although
these exposures tend to have short exposure times, from 20s to 200s,
the large field-of-view of CAFOS CCD
($\sim$15$\arcmin$$\times$15$\arcmin$) and pixel size
(0.53$\arcsec$$\times$0.53$\arcsec$), allow a good estimation of the
sky background. The use of calibration fields, with more than 4
calibration stars observed per field, guarantees a good estimation of
the magnitude zero-point per filter and exposure (photometric
calibration error $<$0.1 mags). We focused our search in nights
dedicated to a single project, aimed to study the time evolution of
the multiband photometry of supernovae (PI: Dr. W. Hillebrandt;
Pastorello et al. 2007), observed in service mode. The reason for that
was that this program uses the same instrument setup, filters,
calibration fields and covers a large periods of time. The number of
nights totally or partially dedicated to this project were $\sim$21
along the considered time period. Finally we selected the darkest
possible nights, ie., moonless nights that distance at least 13 days
from the full moon. We were then left with 5 nights. However, only in
two of them the calibration fields were observed at least one hour
after the begining of the astronomical night. In the remaining 3
nights there was still substantial contamination from the twilight,
and they are therefore excluded from any further analysis. Table
\ref{tab_data} shows the final list of selected nights, including the
broad-band filters observed each night.

The Landolt calibration fields PG1633 and PG0918 (Landlot 1992) were observed
each night, respectively, in the listed filters. The images were reduced
following the standard steps, using IRAF routines. First a master bias frame
was created for each night by averaging all the bias frames obtained along the
night and smoothing it. All images were then bias corrected by substracting
the corresponding master bias. For each band a master flat-field frame was
obtained by averaging the bias corrected domeflat images, and normalizing to
the median intensity value. Images were then corrected for pixel-to-pixel
response variations dividing by their corresponding flat-field frames. No sky
subtraction was performed since the aim of this study is to determine its
intensity.

\subsection{Extinction data}
\label{ext_data}

The Calar Alto Extinction monitor (CAVEX), is an inhouse developed instrument
(PI: U.Thiele), which estimates the monocromatic extinction in the V-band
continously along each night. The system is fully automatic, opening half an
hour after the beggining of the astronomical night and closing half an hour
before its end. It continously points towards the north (the polar star),
taking images of $\sim$20s exposure time every $\sim$76s covering a
field-of-view of $\sim$55 degrees with a resolution of 2.3$\arcmin$/pixel. By
traking the location of 15-20 stars in the field, it estimates the extinction
by measuring their apparent magnitudes across a range of 1.1-2.4 airmasses. It
shares the same limits of humidity and wind speed than the telescopes at the
obsevatory, producing a measurement of the extinction every $\sim$2 minutes if
the night is clear. When the night is cloudy or the extinction has strong
fluctuations, the instrument does not produce a reliable data, flagging
it. Therefore, the fraction of time without precise extinction measurements
from the CAVEX is a good estimation of the amount of time lost due non
astronomical weather conditions in the observatory.  We have collected all the
available data from the extinction monitor during the last 4 years, from May
2003 to May 2007. The data comprises 214193 individual measurements,
corresponding to 1044 nights from a total of 1478 nights included in this
period.

CAVEX estimates the total extinction in a single band. To characterize
the extinction curve it is needed to measure the extinction at
different wavelengths. In the Summer of 2006 (27/07/06-15/08/06) and
the Winter of 2006-2007 (17/12/06-13/01/07), an instrument built with
the sole purpose of estimating this extinction curve was installed,
named the Extinction Camera and Luminance Background Register
(EXCALIBUR, PI: J.Aceituno).  EXCALIBUR is a robotic extinction
monitor able to do cuasi-simultaneous photometric observations in 8
narrow bands covering the wavelength range between 3400\AA\ and
10200\AA, characterizing the extinction curve in this wavelength range
(although only 6 of the bands were operative when installed at Calar
Alto). The instrument was built to estimate the aerosol abundance
based on the shape of the extinction curve (P\'erez-R\'amirez
2007a,b). It estimates $\sim$16 extinction coefficients for each of
the sampled bands per hour, and an average of $\sim$160 extinction
coefficients per band for each night. It was operative for a total of
6 nights in the Summer season and 14 nights in the Winter season. We
collected all these data for the current study.

\subsection{Seeing data}
\label{seeing_data}

The night-sky seeing is measured by a Differential Image Motion
Monitor (DIMM, Aceituno 2004) at the Calar Alto observatory since
August 2004, and it is nowadays a fully automatic instrument. It measure the
seeing at the wavelengths corresponding to the Johnson $V$-band. This
DIMM was calibrated with a previous one installed in the observatory
\cite{vern95}, which, at the same time, was also calibrated with a
Generalized Seeing Monitor in May 2002 \cite{ziad05}, when they both
estimated a median seeing of 0.90$\arcsec$ for that time period. To
study the possible evolution of the seeing at the observatory and its
possible dependencies with another night-sky parameters we collected
all the available seeing measurements on the time period between
January 2005 and December 2006. Contrary to the CAVEX, the DIMM has
more severe limitations to be operative, and it closes automatically
when the humidity is larger than an 80\% or the wind speed is 12 m
s$^{-1}$, which comprises $\sim$50\% of the time.  The DIMM produces
an estimation of the transversal and horizontal seeing in each
measurements. Whenever there is an occasional large difference between
both estimations (not due to a dominant wind component) they most
probably are not due to atmospheric effects, but rather to mechanical
oscilations. Due to inherent stability of a DIMM system to this kind
of oscilations, these cases comprise a relatively small number
($\sim$5\%). They have been excluded from the final dataset.  We
finally collected a total of 213622  seeing measurements 
distributed along 335 nights during the considered time period.

\section{Analysis and Results}
\label{ana}

We describe here the analysis performed over each of the different collected
data. 

\subsection{Night Sky Spectrum}
\label{ana_spec}

\begin{figure}
\resizebox{\hsize}{!}
{\includegraphics[width=\hsize,angle=270]{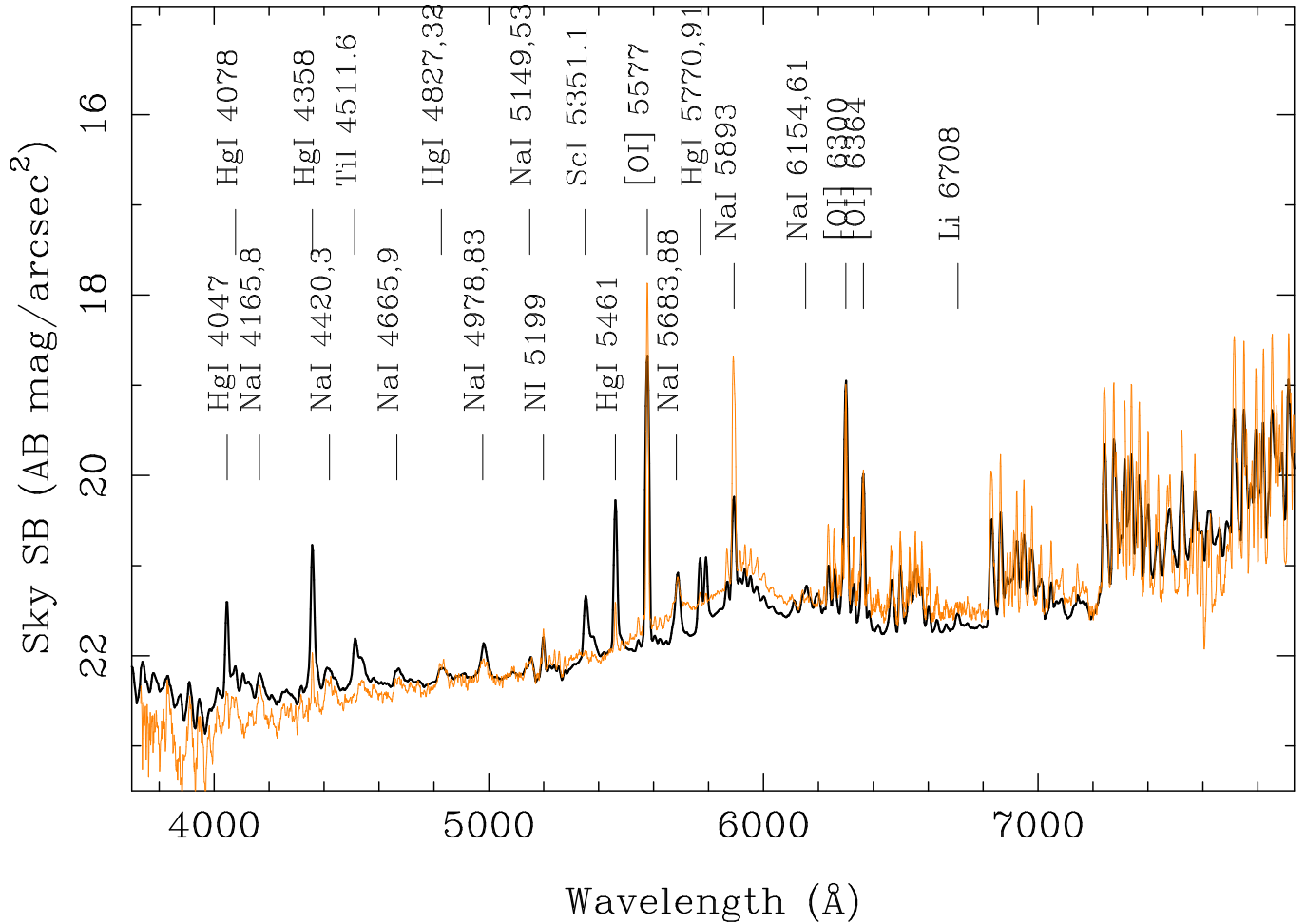}}
  \caption{\label{spec} 
    Night-sky spectrum at the Calar Alto Observatory in the optical wavelength
    range (3700-7950\AA), obtained after averaging 10 spectra of 6 moonless
    nights pointing near the zenit (Black solid-line). The intensity has been
    scaled to that of the darkest moonless night in the $V$-band. Several
    emission lines are indentified in the spectrum. The most relevant ones
    have been labeled with its corresponding name and wavelength. In addition,
    the broad-emission band of NaI centred at $\sim$5900\AA, and the water
    vapor Meinel bands are clearly identified in the spectrum. For comparison
    purposes we included the night sky spectrum at the Kitt Peak observatory
    derived by Massey \& Foltz (2000), obtained from their webpage: 
    {\it http://www.lowell.edu/users/massey/nightsky.html} (Orange dotted-line). It is
    appreciated how  strong are the pollution lines at Calar Alto, in
    comparison with that observatory. 
 }
  \end{figure}

The data included in the final sample of night-sky spectra for
moonless dark nights described in Section \ref{spec_data} were
combined to create a typical moonless night-sky spectrum at the
observatory. For doing so each spectrum was normalized to the mean
intensity (of the sample) at 5200\AA\ (0.13 \funits). Then the final
spectrum was produced by averaging the flux of these normalized
spectra at each sampled wavelength. Figure \ref{spec} shows the
resulting spectrum, covering a wavelength range between 3700 and
7933\AA. This is the first time that a typical night-sky spectrum is
published for the Calar Alto observatory \footnote{The final reduced
  FITS file can be downloaded from the webpage:
  http://www.caha.es/sanchez/sky/}. The emission lines clearly
identified in the spectrum have been labeled with its corresponding
atomic name and wavelength.  Several of the distinctive features of
the night-sky spectrum are due to airglow, although a substantial
fraction is due to light-pollution.

The airglow is emitted by atoms and molecules in the upper atmosphere which
are excited by solar UV radiation during the day and twilight (Ingham 1972).
Airglow is the most important component of the light of night-sky. It
produces the 5577\AA\ and 6300\AA\ lines from OI (which are stronger near the
twilight). It contributes to the ubiquitious 5890-6\AA\ NaD doublet, although
this line is heavely contaminated by light-pollution from low and high
preassure street-lamps. Airglow is also responsible for the OH
rotation/vibration bands in the red and IR, known as the Meinel bands (Meinel
1950), visible at the redder wavelengths of the spectrum (see Fig.
\ref{spec}). In addition it produces a pseudo-continuum in the blue from
overlapping O$_2$ bands (2600-3800\AA) and in the green from NO$_2$ bands
(5000-6000\AA). A more detailed description of the effects of the airglow on
the night-sky emission can be found in Benn \& Ellison (1998a). An atlas of the
airglow from 3100\AA\ to 10000\AA\ was presented by Ingham (1962) and
Broadfoot \& Kendell (1968).

Other contributions to the night-sky spectrum are the zodiacal light,
the starlight, and the extragalactic light, which increases the
background continuum emission (see Benn \& Ellison 1998a and
references therein for a more detailed explanation on their
effects). All of them, including the airglow, comprises the natural
processes that produce the night-sky spectrum in any astronomical
site. In addition to them, the night-sky spectrum can be affected by
the light-pollution due mostly to the street-lights of populated areas
near the observatories. Light pollution arises principally from
troposheric scattering of light emitted by sodium and mercury-vapour
and incandescent street lamps (McNally 1994; Holmes 1997).

Typical spectra of the three more used types of street-lamps were presented by
Osterbrock et al. (1976): the sodium low-pressure lamps, the sodium
high-pressure lamps and the mercury lamps.  The first of the three is the one
with less impact in astronomical observations, since its produces most of its
light concentrated in the 5890-6\AA\ NaD and 8183-95\AA\ NaD emission
lines. Therefore only a very reduced number of astronomical observations are
strongly affected by them. On the other hand the high-pressure sodium lamps
emitt most of its light in a broad NaD line centred in $\sim$5890\AA, with a
FWHM of $\sim$400\AA, that shows a central reversal. They also show a strong
8183-95\AA\ emission line, and fainter emission lines at 4494-8\AA, 4665-9\AA,
4758-52\AA, 4979-83\AA, 5149-53\AA, 5683-88\AA, 6154-61\AA\ and
7665,7699\AA. The high-pressure sodium lamps may affect strongly the quality
of any astronomical observation. Finally the mercury lamps produce narrow
lines at 3651/63\AA, 4047\AA, 4358\AA, 5461\AA, 5770\AA\ and 5791\AA, together
with broad features at 6200\AA\ and 7200\AA\ of FWHM$\sim$100\AA, from the
phosphor used to convert UV to visible light. They also produce a weak
continuum emission over the whole visible range. Some mercury pollution lines
can strongly affect certain astronomical studies: (i) the 4358\AA\ Hg line
strongly affect any attempt to measure the emission of the 4364\AA\ [OIII]
line in any object in the Galaxy. This line is fundamental for the estimation
of the electron temperature of galactic nebulae. (ii) the 5460\AA\ Hg line
lies in the centre of the $y$-band of the $ubvy$ Str\"omgren photometric
system, which may affect the programs devoted to the study of stellar
populations that use this filter set.

Other sources of light pollution are the incandescent lamps and the
high-pressure metal halide lamps. The spectrum of the former consists of
continuum emission only, and it is difficult to identify in a night-spectrum.
The latter are nowadays frequently used in the illumination of sport stadiums
and arquitectonical monuments (which can be considerable, since 
they are normally oriented towards the sky). These high-pressure metal halide
lamps exhibit some Scandium, Titanium and Litium emission lines, that are
charactezed by a blue edge due to molecular bands (General Electric 1975; Lane
\& Garrison 1978; Osterbrock et al. 1976).

The night-sky spectrum shown in Fig.\ref{spec} shows clear evidence of
strong light pollution from all the street-lamps described before. It
shows strong Mercury lines all over the spectrum, the typical emission
lines and features of high-pressure sodium lamps, with a well detected
broad emission at $\sim$5900\AA, and a strong NaI emission line at
5893\AA\ indicative of low-pressure sodium lamps. Evenmore, it also
presents the typical emission lines corresponding to high-pressure
metal halide lamps. All this pollution comes from the populations
nearby the observatory, in particular from Almeria ($\sim$250000
habitants), at 40 km towards the south, and smaller towns at the north
(like Baza, $\sim$21000 habitants, and Macael, $\sim$6000
habitants). There are well established estimations of the contribution
of city lighting to dark-sky brightness (e.g., Treanor 1973, Walker
1973, Yocke et al. 1986, Garstang 1991, and reference
therein). However, its contribution to the spectrum is more complex,
since it depends on the particular kind of lamps used for street
illumination. Most major observatories nearby populated areas have
some kind of night-sky protection laws with the aim of reducing the
effects of light pollution by controlling the fraction of light that
escapes towards the sky and the kind of lamps used (normally they
promote the use of low-pressure sodium lamps). The Calar Alto
Observatory does not benefit yet from any local sky-protection law
which regulates the street illumination, with the corresponding
effects that can be appreciated in the night-sky spectrum.

To estimate the contribution of light pollution to the night-sky spectrum we
derive the flux intensity corresponding to each of the detected emission
lines. Each line was fitted with a single gaussian function using FIT3D
\cite{sanc06b}, using the same procedure described in \cite{sanc07b}.  Table
\ref{tab_line} lists the integrated flux for each of the detected emission
lines shown in Fig.\ref{spec}, together with the identification of the line
and the nominal wavelength. In addition to the emission lines, the flux
intensity of the sodium broad-band emission at $\sim$5900\AA\ was also
estimated by fitting the feature with a single broad gaussian function. The
result is also listed in Table \ref{tab_line}. All the fluxes were converted
to Rayleighs following the conversion formulae by Benn \& Ellison (1998a).

These values can be compared to those ones obtained for another
astronomical sites. E.g., Pedani (2005) presented a study of the
night-sky spectrum at the Observatory of the Roque de los Muchachos,
La Palma. The intensity of the lines produced by the Mercury and
low-pressure Sodium lamps found in our spectrum are only comparable to
those ones in La Palma when pointing in the directions towards the
most polluting towns in this island. On the other hand, the
contribution of the lines produced by the high-pressure Sodium lamps
is much lower, being comparable to that of the less polluted areas of
the sky in La Palma. This may indicate that most of the street lamps
used in Almeria are Mercury and low-pressure Sodium lamps, rather than
high-pressure Sodium ones, and therefore they only affect specific
wavelength ranges (and science programs). Finally we clearly see lines
produced by high pressure metal halide lamps, only marginally (and
recently) detected in the sky spectrum at La Palma (Pedani 2005).

Once determined the contribution of each emission line to the
night-sky spectrum it is possible to decontaminate it by the effect of
the pollution lines (ie., all but the OI ones), creating a {\it clean}
night-sky spectrum. This can be done directly for the Mercury,
Scandium, Titanium and Litium lines, since all its emission is
produced by light pollution. However, in the case of the Sodium lines
there is a natural contribution due to the airglow. We lack a direct
measurement of this natural contribution at Calar Alto, that can only
be achived nowadays with a general blackout. However its natural
contribution must not be significantly larger than in other
astronomical sites. Benn \& Ellison (1998a) estimated the natural
contribution by the broad Sodium emission band in $\sim$0.03 mag in V
and $\sim$0.02 mag in R for La Palma observatory. Similar
contributions are expected by the NaI 5893,6 emission
lines. Therefore, the natural contribution by the Sodium due to the
airglow is expected to be of the order of $\sim$0.04 mag in both
bands. Thus, as a first order we can assume that all the detected
emission by Sodium in our sky-emission spectrum is due to pollution,
and once determined its contribution we can correct it by the expected
contribution of natural emission, if needed.

Table \ref{tab_cont} lists the estimated contribution of the light pollution
to the sky background in different bands. It includes the $B$, $V$ and
$R$-Johnson filters, and a set of medium band filters selected from those ones
used by the ALHAMBRA survey (Moles et al. 2007), a major imaging survey
currently on going at the Calar Alto observatory. They are included to
illustrate the effects of the pollution lines in the background sky when using
median/narrow-band filters affected by these lines. The contamination from
light pollution is clearly stronger in the $B$-band than in other astronomical
sites: e.g, $\sim$0.02 mag at La Palma (Benn \& Ellison 1998a), 0.02-0.04 mag
at Kitt Peak (Massey et al. 1990). Although there are astronomical sites with
similar or stronger contaminations: e.g. Mount Hopkins (see Fig. 2, from
Massey \& Foltz 2000). This contamination can be reduced by a proper light
pollution law that limits the use of the Mercury street lamps. Strong benefits
from such laws has been experienced in different sites (e.g., Benn \& Ellison
1998a; Massey \& Foltz 2000). The contamination in the $V$ and $R$-bands is
slightly stronger than in some other places, like La Palma: 0.05-0.10 mag at the
$V$-band and 0.07-0.12 mag at the $R$-band (Pedani et al. 2005), but it is
similar or smaller than in other astronomical sites, like Kitt Peak: 0.19-0.33
mag in the $V$-band (Massey et al. 1990) or Mount Hopkings: 0.17 mag in the
$V$-band (Massey \& Foltz 2000). One of the two recomendations of the IAU for
a dark place is that the contribution of the pollution to the Sodium emission
should not exceed in intensity the natural airglow one (Smith 1979). If we
consider that the airglow emission is similar in Calar Alto than in La Palma
(or at least of the same order), it is clear that the contribution from the
light pollution is much stronger. In these regards Calar Alto does not
fulfill the IAU recomendations for a dark place. A proper light-pollution law,
that increases the use of low-preasure sodium lamps rather than high-presure
ones, will not reduce the net effect of the light pollution to the
sky-background in these bands, but it will concentrate it in a more reduced
wavelength range, affecting less observing programs.

Most of the lamps that cause the light pollution also produce a certain level
of continuum emission. In particular Mercury and high pressure metal halide
ones. However its contribution is difficult to estimate. Although we donecannot
quantify its effect, the reduction of the use of Mercury, high-pressure Sodium
lamps and high pressure metal halide lamps in the vicinity of the observatory
will certainly also reduce the sky-background continuum.

The lack of previous similar studies of the night sky spectrum at the Calar
Alto observatory (to our knowledge) does not let us to analyze the evolution
of the light pollution along the time.

\subsection{Night Sky Brightness}
\label{ana_mag}

The night sky brightness was determined by using both the imaging and
spectrocopic data described before. In the case of the imaging data, the
calibration fields contain at least 5 calibration starts per frame. The
magnitude zeropoint for each image was determined by measuring the counts
intensity of each of these stars within a fixed aperture of 8$\arcsec$ radius,
using IMEXAM task of IRAF package, and applying the formula:

$$mag_{zero} = mag_{app} + ext + 2.5 {\rm log}_{10} ( counts/t_{exp}) $$

where $mag_{zero}$ is the magnitude zeropoint, $mag_{app}$ is the apparent
magnitude of the calibration star in the corresponding band, $ext$ is the
extinction, derived from the correponding value measured by the CAVEX this
night and the airmass of the image, $counts$ are the measured counts within
the indicated aperture, and $t_{exp}$ is the exposure time. Since the measured
seeing of the images (FWHM of the field stars), ranges between 0.9$\arcsec$
and 1.3$\arcsec$, this aperture ensures that most of the flux is contained
within it, and no aperture correction was applied. The average of the values
derived for each calibration star in the field is considered the zeropoint of
the image, and the standard deviation from this mean value is considered as
the photometric error. This standard deviation ranges between 0.02 mag and
0.06 mag for each band and each night.

The sky brightness was then determined in each image by measuring the mean
counts level in several ($>10$) square apertures of
$\sim$50$\arcsec$$\times$50$\arcsec$, in areas free of targets, using IMEXAM
task of IRAF package. The mean value of these measurements is considered as
the count level of the sky brightness, and the standard deviation with respect
to this mean value the error in the count level estimation. Finally, the sky
surface brightness was determined using the formula:

$$SB_{sky} = mag_{zero} - 2.5 {\rm log}_{10} ( counts_{sky}/t_{exp}/scale^2) $$

where $SB_{sky}$ is the surface brightness in magnitudes per square arcsec,
$mag_{zero}$ is the zeropoint described before, $counts_{sky}$ is the sky
counts level estimated and $scale$ is the pixel scale in arcseconds (ie.,
0.53$\arcsec$ for CAFOS).  It is noticed that the magnitude zeropoint was
corrected by the extinction, but the sky brightness was not, following the
convention adopted in most of the recent studies of sky brightness (e.g.,
Walker 1988b; Krisciunas 1990; Lockwood et al 1990; Leinert et al. 1995;
Mattila et al. 1996; Benn \& Ellison 1998a,b). Correcting the sky brightness
for extinction would be appropiate only if the extinguishing layer were known
to be below all sources of sky brightness, which is not the case (Benn \&
Ellison 1998a,b).

In addition to the estimations of the sky surface brighness obtained using
direct imaging we also derived the sky surface brightness by using the PPAK
spectroscopic information for the only full moonless night of our sample
(02/06/2005). The two night-sky spectra obtained this night cover the
wavelength range of the $B$, $V$ and $R$-band filters. We derived the sky flux
intensity for each of these filters by convolving each spectrum with the
corresponding filter tranmission curves listed in the Asiago Database on
Photometric Systems (Moro \& Manuri 2000). Then the fluxes were transformed to
magnitudes by using the zero-pointings listed in Fukugita et al. (1995). The
mean value of the two derived magnitudes in each filter is adopted as the sky
surface brightness of that night, and the absolute difference between both
magnitudes as the error.

Table \ref{tab_mag} lists the sky-surface brightness obtained from both the
imaging and spectroscopic data at the different bands for each moonless
night. In addition, it lists previous results on the sky-brightness at
different astronomical sites, including the results presented by Leinert et
al. (1995) for Calar Alto, derived from broad-band photometry obtained along
three nights in 1990.

\subsubsection{Dependency on the zenith distance}

There is a wide range of sky surface brightness values for the different
bands, considering that all the measurements were obtained in full moonless
nights, without twilight contamination. The data from the first night, derived
using PPAK data, are very similar to those ones from the third night, derived
using CAFOS data, being the sky in the former slightly brigher in the
$B$-band. The sky brightness of the second night is brighter in all the
bands. Looking back to the data, we realize that the airmass of the images of
the 1st night ($\sim$1.7), corresponds to an elevetion much lower than that of
the 2nd and 3rd night data ($\sim$1.3 and $\sim$1.2, respectively). The sky
brightness increases with the airmass for two different reasons. One is a
natural effect of the airglow, which is brighter at low elevations cause the
line of sight intercepts a larger number of atoms in the airglow layer
(Garstang 1989; Benn \& Ellison 1998a, and references therein). A second
effect is the increase of light pollution when pointing towards high populated
areas at low airmass. Walker (1971,1991) and Garstang (1989) estimate that
the increase in the brightness at an air mass of $\sim$1.4, in the direction
of a populated area of P inhabitants at a distance D km to be $\sim \frac{P
D^{-2.5}}{70} $ mag. That would correspond to $\sim$0.3 mag when pointing
directly towards the south, where is the largest populated city nearby
(Almeria). However since our observed fields are mostly pointing towards the
east its contribution is more difficult to estimate. It is not possible to
know which is the actual contribution of the light pollution to the continuum
brightness at the zenith. Therefore we do not know if Calar Alto fulfill (or
not) the other IAU recomendation for a dark site, that is that this
contribution must be lower than $\sim$0.1 mag (Smith 1979).

It is possible to derive an approximate expression of the sky brightness
dependecy on the zenith distance due to natural effects based on the results
by Garstang (1989), as already pointed out by Krisciunas \& Schaefer
(1991). This expression can be used to correct the measured values and derive
a much appropiate estimation of the zenithal sky brightness (Benn \& Ellison
1998a; Patat 2003). Patat (2003) derive the following formula (Appendix C of
that article) for this correction:

$$\Delta m = -2.5 {\rm log}_{10}[(1-f)+f X]+\kappa (X-1)$$

where $\Delta m$ is the increase in sky brightness at a certain band and
airmass ($X$), $f$ is the fraction of the total sky brightness generated by
airglow, being (1-$f$) the fraction produced outside the atmosphere (hence
including zodiacal light, faint stars and galaxies) and $\kappa$ is the
extinction coefficient at the corresponding wavelength.

We applied this correction to our data, using the typical extinction curve at
Calar Alto (following sections) normalized to the corresponding $\kappa_V$
extinction coefficient of each night.  A typical value of $f=$0.6 was used for
this correction (Patat 2003).  Once applied there is a significant reduction
of the dispersion between the values obtained for each night. This indicates
that the correction works pretty well, despite the fact that it does not take
into account the effects of light pollution. The mean values of the sky
brightness at the zenit after correction for each band are also listed in
Table \ref{tab_mag}.

\subsubsection{Variation of the sky-brightness along the time}

The night sky-brightness at Calar Alto shows no significant change in the last
15 years, when comparing with the results by Leinert et al. (1995). They did
not applied any correction for the dependecy on the zenith distance to their
data (Table 6 of that article). Therefore we must compare with the mean values
without correction. The only band where it is appreciated an increase of the
sky-brightness is the $U$-band ($\sim$0.4 mag brigther). However, if we take
into account that we only have data for this band at low elevation, we cannot
consider these results conclusive. For the remaining bands the sky seems to be
$\sim$0.2 mag brighter in the $B$-band, $\sim$0.2 mag fainter in the $R$-band
and without changes in the $V$- and $I$-bands, when comparing with the mean
values derived for the three nights of our sample. However none of these
differences seems to be significative, lying withing the errors of our
measurements. When comparing with the two darkest nights with data obtained at
high elevation the differences (if any) disappear for the $B$ and $V$-bands,
and the sky seems to be even darker in the $R$ and $I$-band. A possible caveat
to this comparison is that the results listed in Leinert et al. (1995) were
obtained not exactly in the solar minimun, which may produce an increase of
the sky-brightness. However, although their broad-band data were obtained in
the 1990, they also obtained intermediate-band data in the 1993 and their
sinthetized broad-band surface sky-brightness are similar to those obtained in
the nights of the 1990.

\subsubsection{Comparison with other astronomical sites}

The sky surface brightnesses at the Calar Alto observatory at
different bands, listed in Table \ref{tab_mag}, are remarkable similar
to those ones at many other different astronomical sites. In the
optical wavelength range ($U$,$B$,$V$ and $R$-bands), Calar Alto seems
to be a particular dark site, comparible to Mauna Kea. The fact that
both places seem darker than Paranal may be an artifact since the data
presented by Patat et al. (2003) were taken during the maximun of
solar activity. This result is in anycase remarkable, considering the
strong light pollution present in the Calar Alto spectra, which effect
is particularly strong in the $B$, $V$ and $R$-bands
(Tab. \ref{tab_cont}). Most of the listed observatories have little
light pollution or they benefit of specific protection laws against
it. This has been demonstrated as a tremendous useful tool to preserve
or increase the quality of the night-sky (eg., Benn \& Ellison
1998a,b; Massey \& Foltz 2000; Walker \& Schwarz 2007). If the effects
of light pollution could be reduced in the vicinity of Calar Alto it
would become a particularly dark site for optical observations.

On the other hand, the sky is clearly brighter in the $I$-band than in any
other astronomical site listed in this table. Despite the fact that the
observatory is located in the most arid place in Europe, in the vecinity of a
desert (the Tabernas desert), the water vapor Meinel bands are particularly
strong. The humidity at Calar Alto is higher than in other astronomical sites,
like Paranal or Mauna Kea, although there are frequent epochs of low humidity
($<$20\%) in the Summer. The height of the observatory, $\sim$2200 m over the
sea level, normally places it under the inversion layer, which has a particular
strong impact in the strength of the water vapor emission lines. Both combined
effects can explain the rise of the sky-brightness in the $I$-band. It is
important to note here that this effect has a relatively reduced effect over
near-infrared observations in the $J$, $H$ and $K$-band.

\subsection{Extinction coeffcients}
\label{ana_dust}

The median $V$-band extinction at the Calar Alto observatory for the time
period covered by the CAVEX data (Section \ref{ext_data}) was $\sim$0.18 mag,
with a mean value of $\sim$0.21$\pm$0.08 mag. This value is slightly smaller
than the previously reported by Hopp \& Fernandez (2002), which was based on a
much smaller sample of data (comprising 74 nights spanned between 1986 and
2000).  They found that there was an increase of the extinction at the Summer
season, that was most probably associated with an increase of the aerosols
(ie. dust) in this period of the year. Similar seasonal pattern has been
appreciated in another major observatories: e.g., La Palma observatory is
strongly affected by dust extinction in the Summer when dust from the Sahara
desert ($\sim$400 km away) blows over the Canary Islands (Benn \& Ellison
1998a). Although the Calar Alto observatory is nearer to the Sahara desert
($\sim$250 km away) than La Palma, it is normally out of its main wind
streams, being shielded by the Altas mountains. On the other hand, it is
located in an arid region nearby a much smaller desert, the Tabernas desert
($\sim$15 km away).

\begin{figure}
\resizebox{\hsize}{!}
{\includegraphics[width=\hsize,angle=270]{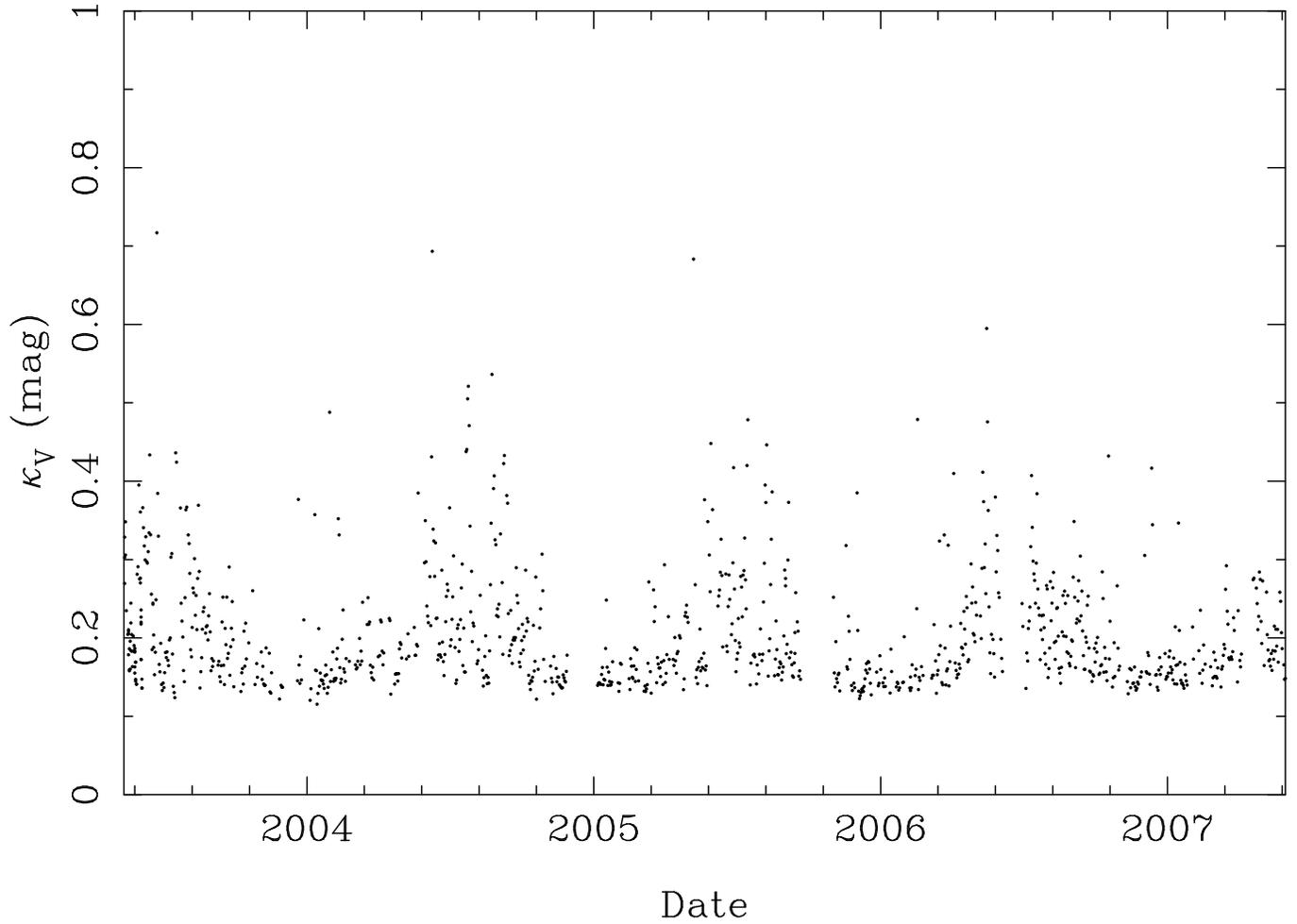}}
  \caption{\label{cavex} 
Distribution of the V-band extinction along 1044 astronomical
useful nights in the period between march 2003 and may 2007 ($\sim$70\% of the
total nights). The extinction shows a clear seasonal pattern, with a maximum
in Summer and a minimun in Winter. The median value of the extinction is
$\kappa_V$=0.18 mags, with a typical value of $\sim$0.15 mags in Winter. The fraction
of nights with $\kappa_V>$0.25 mags, $\sim$20\%, is similar to other observatories
(e.g., La Palma), although the peaks of extinction are much lower.
}
  \end{figure}

Figure \ref{cavex} shows the evolution of the average $V$-band extinction for
each night along the period of time sampled by the dataset. As already
suspected by Hopp \& Fernandez (2002), there is a clear seasonal pattern. The
typical extinction in the Winter nights is $\kappa_V\sim$0.15 mag, being
mostly restricted to values lower than $\kappa_V<$0.2 mag. In Summer time
there is a wider range of extinctions, although in most of the cases the
extinction is lower than $\kappa_V<$0.4 mag. As indicated before this increase
of the extinction is most probably associated with a rise of the aerosols
(dust) in the atmosphere. We will explore that possibility latter.

This seasonal pattern is somehow similar to the one seen in La Palma. Indeed
the fraction of nights with $\kappa_V>$0.25 mag is similar in both
observatories, $\sim$20\% of the nights. However there is a major difference:
the fraction of nights with high extinction, $\kappa_V>$0.4 mag, at Calar Alto
is very reduced, $\sim$3\%, while at La Palma this fraction is $\sim$10\% of
the nights, with frequent peaks of extinction over $\kappa_V>$0.6 mag  (Benn
\& Ellison 1998a, Figure 3).

Based on the fraction of nights that the CAVEX was operative and derived
realiable measurements of the $V$-band extinction ($\kappa_V$), it is estimated
that $\sim$70\% of the nights were astronomically useful in the period covered
by these data (4 complete years). This fraction is remarkable similar to that
one in many other astronomical sites (eg., La Palma, Benn \& Ellison 1998a).
The fraction of fully photometric nights, defined as nights where the $V$-band
extinction varies less than a 20\% along all the night, was $\sim$30\%.

\subsubsection{Contributions to the extinction}

The EXCALIBUR data described in Section \ref{ext_data} were used to determine
the typical extinction curve at Calar Alto. This curve was previously studied
by Hopp \& Fernandez (2002), using an inhomogenous dataset. The extinction
coefficients were analyzed separately for the Summer and Winter seasons due to
the observed seasonal pattern. First, it was derived the mean extinction
coefficients per night by averaging all the measured extinction coefficients
per band obtained along each night ($\sim$160 values). Then, the mean
extinction coefficients per season were determined by averaging all the
measured extinction coefficients per band obtained along each season
nights. Table \ref{tab_ext} lists the average extinctions coefficients
obtained for each season for each of the 6 bands sampled by the instrument,
including their central wavelengths and the standard deviation with respect to
these mean values. As expected the standard deviation in the extinction
coefficients is larger for the Summer data than for the Winter ones, in
agreement with the results shown in the previous section. The EXCALIBUR
results are consistent with the CAVEX ones for the wavelength covered by both
instruments, ie., $\sim$500 nm. This indicates that the extinction
coefficients listed in Table \ref{tab_ext} are a good representation of the
typical values for each season.

The total extinction is mostly due to three contributions, the Rayleigh
scattering at the atmospheric atoms and molecules, the extinction due to
aerosol particles (mostly dust), and extinction due to Ozone (Walker
1987b). The Rayleigh scattering can be described by 

 $$\kappa_{RC} = B(p,t,n') \lambda^{-4}$$

where $\lambda$ is the wavelength and $B$ is a constant that mostly depends on
the pressure, the temperature and the normalized refractive index of the air
($n'$, slightly wavelength dependent). In general $B$ can be replaced by its
average value for the mean conditions in a certain astronomical site (e.g.,
Rufener 1986). The aerosol particles produce a similar absorption, that can be
described by

 $$\kappa_p = b(h_{obs}) \lambda^{-\alpha}$$

where $b$ is a parameter that depends mostly on the height of the observatory
($h_{obs}$), and $\alpha$ is a power law index that depends on the size of the
aerosol grains. Although an $\alpha\sim$1.3, derived by Siedentopf (1948), is
widely used, we adopted a value of $\alpha=$0.8 for consistency with the
previous study of the extinction curve at Calar Alto (Hopp \& Fernandez
2002). The Ozone extinction is a selective absorption by molecular bands. It
can be approximately described by a broad gaussian function centred in
$\lambda\sim$6000 \AA\ (matching the shape shown in Runefer 1986 and Hopp \&
Fernandez 2002):

 $$\kappa_{\rm O3} = { C } \ {\rm exp} [-\frac{\lambda-6000}{1200}]$$

Each single contribution to the extinction depends on the wavelength and a
particular constant that, in the case of the two first, depends on the height
and the average atmospheric conditions in the observatory. We adopted the
values listed for $\lambda =$5400\AA\ in Rufener (1986), consitent with those
derived by T\"ug (1977), obtained for La Silla observatory. A similar approach
was followed by Hopp \& Fernandez (2002). The height and average weather
conditions in this observatory are very similar to those of Calar Alto, which
justifies the use of these constants. 

Therefore the total extinction curve is a linear combination of these three
contributions:

$$\kappa_\lambda = f_1\ \kappa_{RC} + f_2\ \kappa_p + f_3\ \kappa_{O3}$$

The extinction coefficients for the two seasons were fitted to this linear
combination, deriving the relative contribution of each one ($f_i$) to the
total extinction. Table \ref{tab_ecurve} lists the results from this fitting
analysis, including the relative contribution ($f_i$) derived for each
component for each season. For comparison it also includes the same relative
contributions derived for Calar Alto by Hopp \& Fernandez (2002), and their
compilation of similar results for different astronomical sites. 

The contribution of the Rayleigh scattering seems to be rather constant for
the two considered seasons, like the Ozone absorption. On the other hand, the
Aerosol contribution rises considerable in the Summer time, being responsible
of the increase of the extinction in this season, as it was thought. The
estimated contributions of the Rayleigh scattering and the Aerosol extinction
are very similar to the values reported by Hopp \& Fernandez (2002) for the
Winter season, which may indicate that both contributions have not changed
considerably with time (their data corresponds to 1986-2000). Both
contributions are also similar to the ones derived for other major
astronomical sites. On the other hand the contribution of the Ozone absorption
seems to be stronger than in previous measurements and other astronomical
sites. Unfortunally the coverage of EXCALIBUR when mounted at Calar Alto did
not allow to perform an accurate sampling of the wavelength range affected
more strongly by the Ozone absorption, which did not let us to be conclusive
on this respect.

\begin{figure}
\resizebox{\hsize}{!}
{\includegraphics[width=\hsize,angle=270]{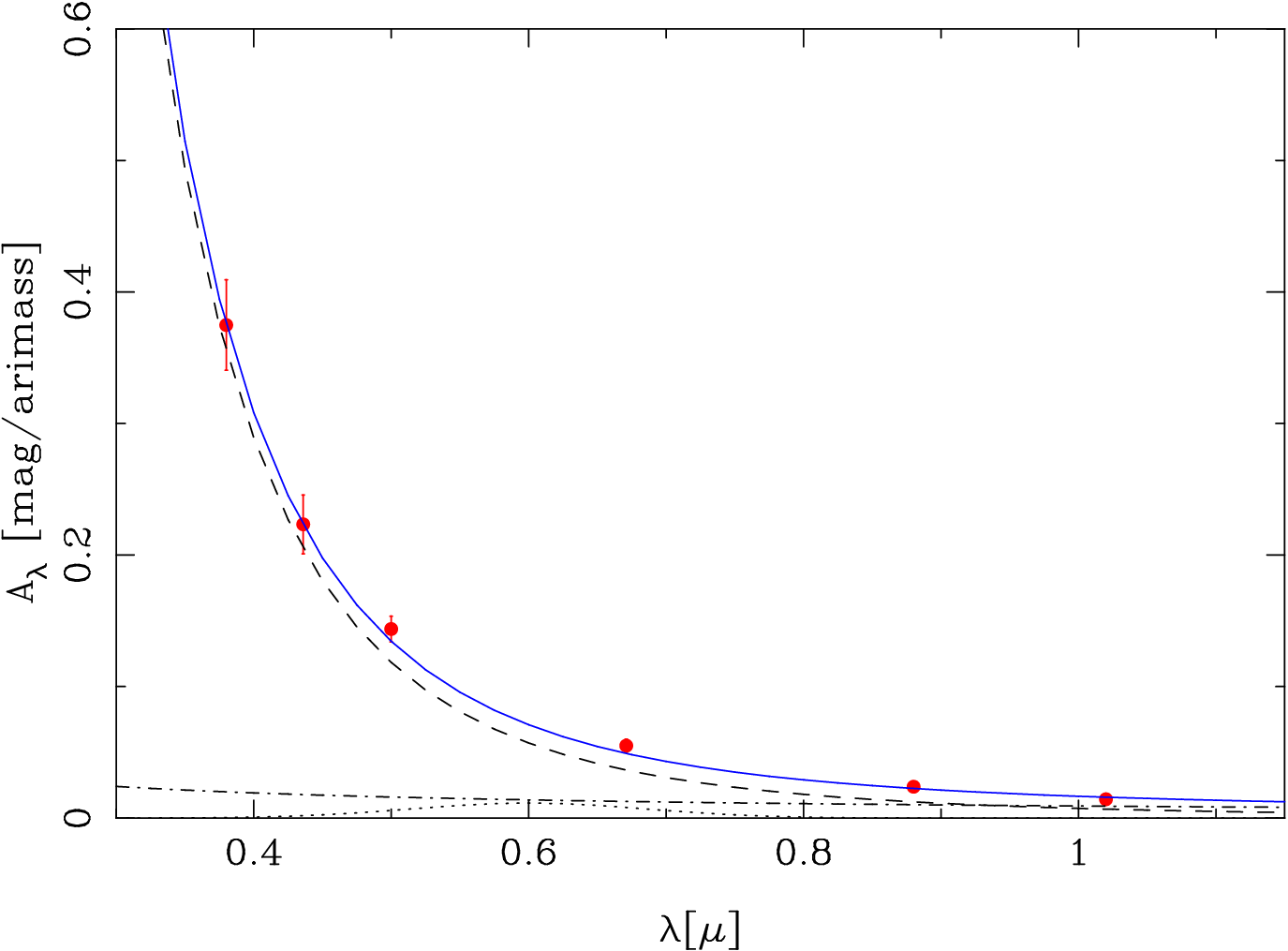}
\includegraphics[width=\hsize,angle=270]{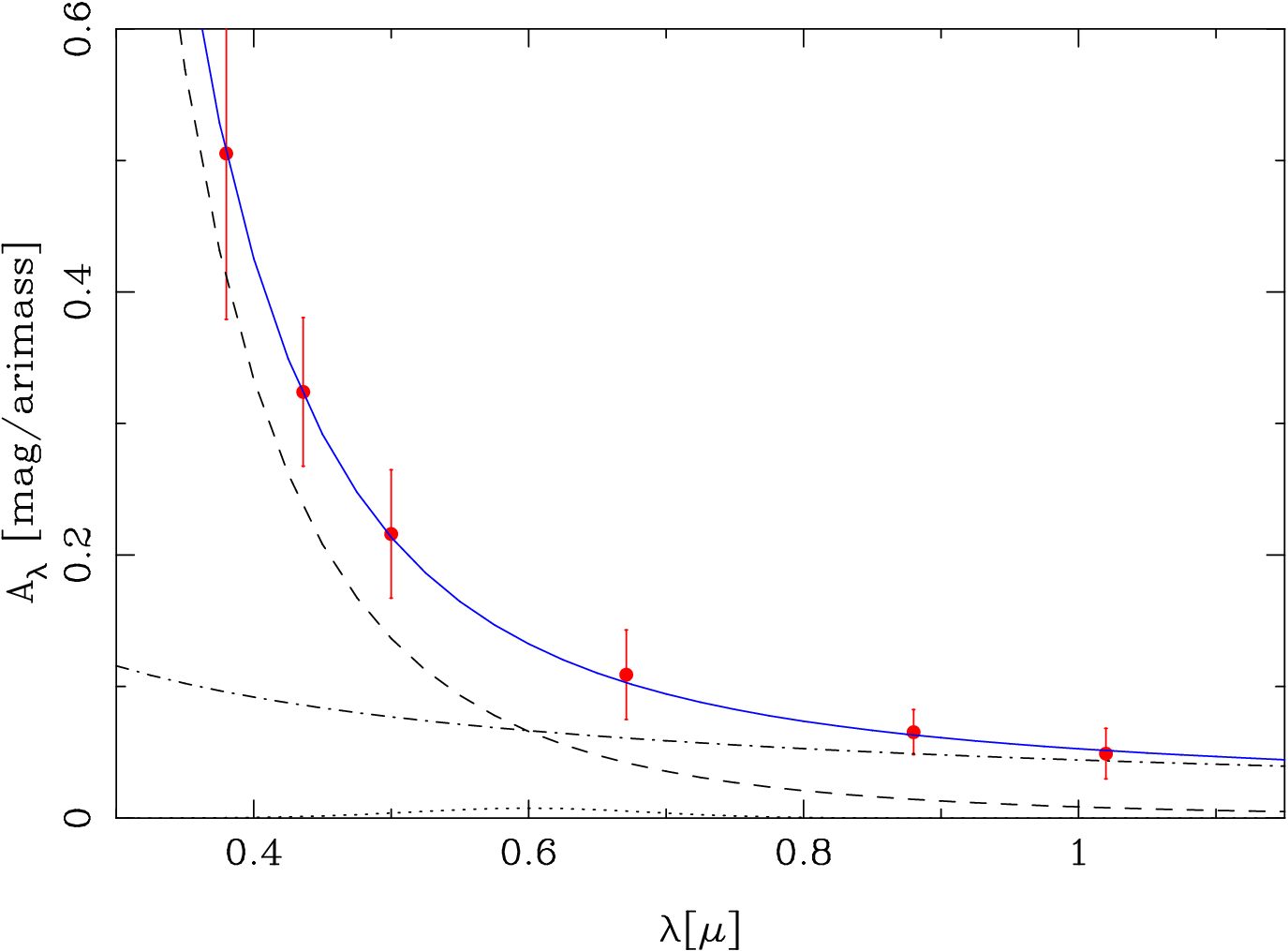}
}
  \caption{\label{ext_curve} 
Extinction curves derived for a set of Winter
(left panel) and summer nights (right panel), using the EXCALIBUR extintion
curve monitor (red solid circles). The error bars indicate the standard
deviation over the mean values.  The data were fitted to a linear combination
of the three different contributions to the extinction described in Walker
(1987): the Rayleigh scattering (dashed line), the Aerosol or dust extinction
(dotted-dashed line), and the Ozone extinction (dotted line), each one scaled
to their relative contribution to the extinction. The blue solid line shows
the best fit to the data.  
}
  \end{figure}

Figure \ref{ext_curve} shows the distribution of the extinction coefficients
along the wavelength for the two season datasets. It also includes the best
fitted linear combination of the three components to the extinction and each
of these components scaled to its relative contribution to the total
extinction. Despite the apparent increase of the Ozone absorption its actual
contribution to the total extinction at any wavelength is very limited, being
neglectible in comparison with the other two contributions. Indeed the fitting
process derives equally good results when this contribution is removed. As
already indicated the Rayleigh scattering contribution is very similar for
both datasets, being the dominant contribution in the Winter season, ie., in
conditions of low extinction. In Winter time is responsible of the $\sim$85\%
of the extinction in the $V$-band, only $\sim$11\% is due to Aerosol
extinction and $\sim$4\% to Ozone absorption. On the other hand, in Summer its
contributions drops to $\sim$63\%, with $\sim$35\% due to Aerosol extinction
and $\sim$2\% due to Ozone absorption. Curiously, the contribution of Aerosols
to the extinction in the conditions of minimun extinction are much reduced
than the one found in other astronomical sites, like La Silla (Burki et
al. 1995).

\subsubsection{The Extinction Curve}

As we shown in the previous section the extinction curve depends strongly on
the relative contribution of each of the three major components to the
extinction. Each of them has a different dependecy with the wavelength, and
two of them depends also on the particular atmospheric conditions at the
observatory. Therefore, it is difficult to derive a precise extinction curve
valid for every night that depends only in a reduced number of parameters.
However it is still possible to look for an approximate expression for the
extinction curve that provide an useful estimation of the extinction at any
wavelength and that depends only in a single parameter: the $V$-band
extinction, measured each night by the CAVEX. 

Based on the results of the previous section it is known that the Ozone
absorption have a marginal effect in the total extintion at any
wavelength. Therefore we have not considered it for our approximate
expression. It is also known that the contribution of the Rayleigh scattering
is almost contant along the year, and therefore the variations in the
extinction are controlled by the amount of Aerosol particles (ie.,
dust). Based on this assumption an approximate expression for the extinction
curve can be derived by considering that the extinction in the $V$-band is due
to a fix contribution of Rayleigh scattering (the average for both seasons)
and a variable contribution due to Aerosol extinction. The derived expression
is:

$$\kappa_\lambda \sim 0.0935\ \left(\frac{\lambda}{5450}\right)^{-4} + (0.8*\kappa_{\rm V}-0.0935)\ \left(\frac{\lambda}{5450}\right)^{-0.8}  $$

The typical differences found in the extinction coefficients derived using
this formula and the more precise decomposition presented in the previous
section are of the order of $\sim$10\%. 

\subsection{Atmospheric Seeing}
\label{ana_seeing}

\begin{figure}
\resizebox{\hsize}{!}
{\includegraphics[width=\hsize,angle=270]{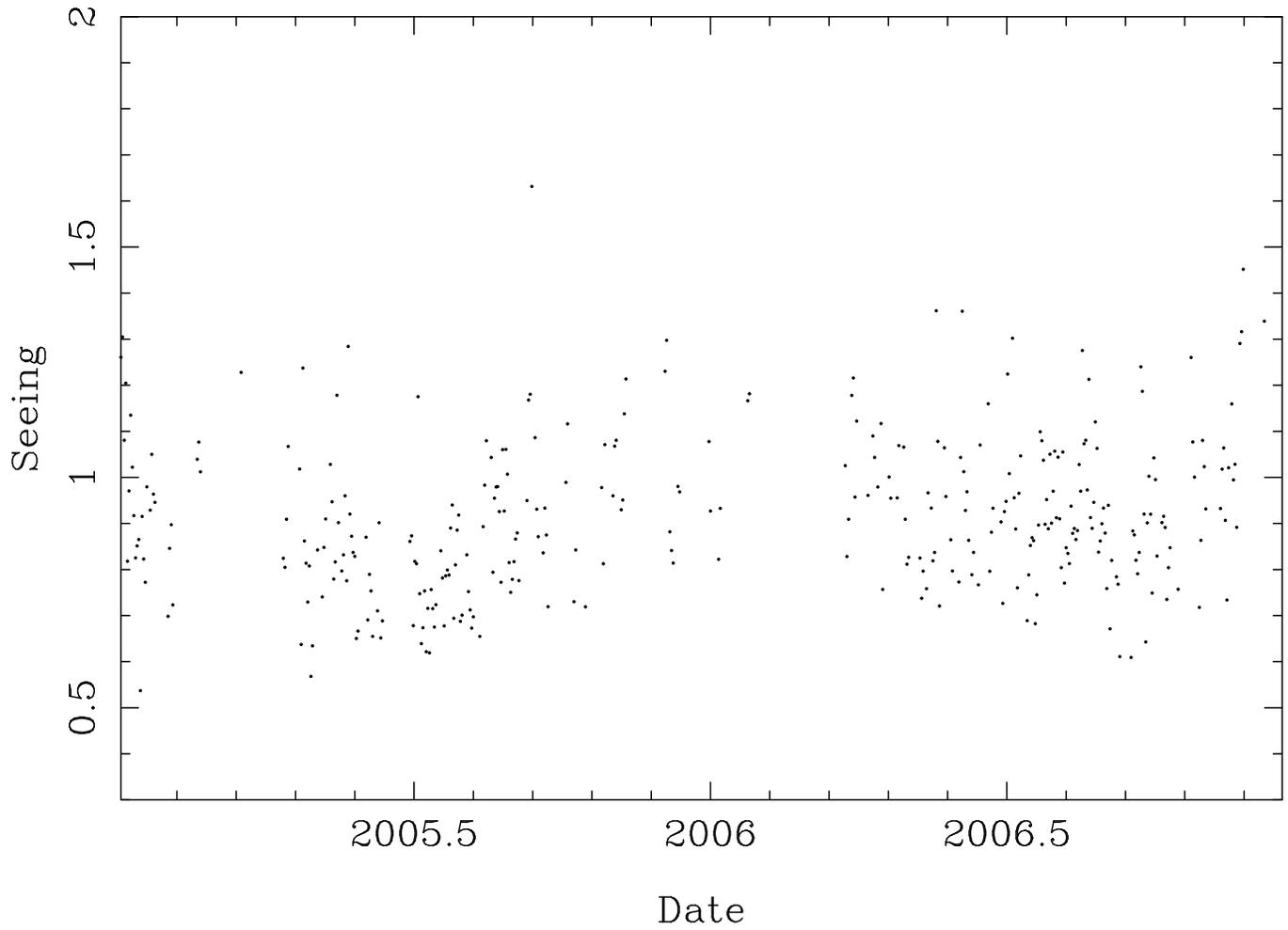}}
  \caption{\label{seeing1} 
Distribution of the seeing along the time for the 335 nights with measurements
from the DIMM monitor in the period between January 2005 and December 2006. 
There is a suggestion for a seasonal pattern, with a minimun during Summer and a 
maximum during Winter. 
  }
  \end{figure}

\begin{figure}
\resizebox{\hsize}{!}
{\includegraphics[width=\hsize,angle=270]{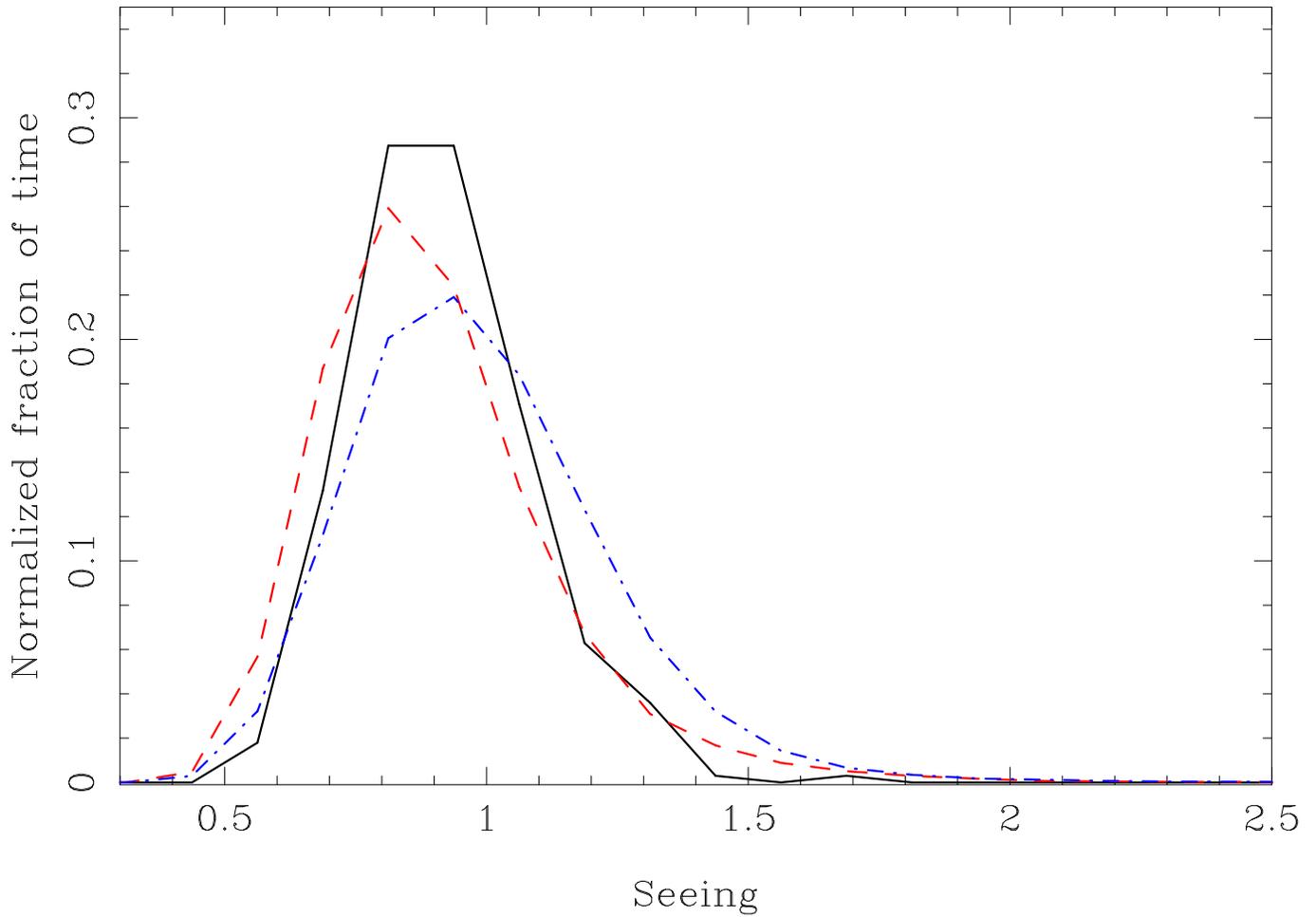}}
  \caption{\label{seeing2} 
Normalized histogram of the seeing distribution shown in Fig.\ref{seeing1}
(solid line). The median value of the seeing is $\sim$0.90$\arcsec$, with a 
$\sim$46\% of the nights with a seeing better than this value. The red dashed
line shows a similar histogram for the summer values (May to September), when
the median seeing drops to $\sim$0.87$\arcsec$. The blue dashed-dot line
shows a similar histogram for the Winter values, comprising the remaining
data, when the median seeing rise to $\sim$0.96$\arcsec$. The seeing
distribution is clearly different in both seasons.
}
  \end{figure}

The seeing data described in section \ref{seeing_data} were used to determine
an average seeing for each night comprised in the dataset (spanned along
$\sim$2 years). The median atmospheric seeing for all the time period was
$\sim$0.90$\arcsec$, with a $\sim$70\% of the nights under subarcsecond seeing
($<$1$\arcsec$). Figure \ref{seeing1} shows the nightly averaged seeing
distribution along the time. The epochs without data in April 2005 and March
2006 were due to reparations in the hut of the DIMM. There is a mild seasonal
pattern in the seeing distribution, with the best seeing concentrated in the
Summer season. To further investigate this possibility we created the seeing
histogram for all the data comprised in the dataset and for two different
subsets of data corresponding to the Summer season (May-September) and the
Winter season (the rest of the months). Figure \ref{seeing2} shows the three
histograms. The differences in the seeing distribution for the Summer (median
seeing $\sim$0.87$\arcsec$) and the Winter seasons (median seeing
$\sim$0.96$\arcsec$) are clearly appreciated. Not only the median seeing is
better in the Summer season, the chances of having better seeing in a Summer
night are larger.

Table \ref{tab_seeing} lists the median seeing estimated for both the total
sample and the two season subsamples. For comparison purposes it also lists
the atmospheric seeing measured at different astronomical sites world-wide,
ordered by increasing seeing. Although the median seeing at Calar Alto is
larger than that of some major astronomical observatories (Mauna Kea, La
Palma), it is actually better than in many other astronomical sites (eg.,
MtGraham, Paranal).

\section{Conclusions}
\label{conc}

We have characterized the main properties of the night-sky at the
Calar Alto observatory, comparing them, when possible, with similar
properties of other different astronomical sites. The main results of
this article can be summarized in the following points:

\begin{itemize}

\item An average night-sky spectrum for the moonless dark-time at the
observatory has been presented for the first time. This spectrum, which covers
the optical wavelength range (3700-7933\AA), is distributed freely to the
community. Airglow and light-pollution emission lines are detected in this
spectrum. The strength of the light-pollution lines has been measured,
estimating their contribution to the emission in different bands. In
comparison with other sites the Mercury lines are particularly strong. The
contribution of the light pollution to the Sodium emission is far stronger
than its natural emission. In this regards Calar Alto does not fulfill the IAU
recomendations for a dark astronomical site (Smith 1979), like other major
astronomical sites (eg., La Palma, Pedani 2005).

\item The moonless night-sky brightness at the zenith has been determined for
  the $U$, $B$, $V$, $R$ and $I$-bands. There was no appreciable change in the
  sky-brightness over the last 15 years. In comparison with other astronomical
  sites, Calar Alto shows a particularly dark sky in the optical bands,
  similar to that of MtGraham or Mauna Kea. The sky brightness could be even
  darker if it was possible to reduce the light pollution in the optical
  bands, which would place Calar Alto as a very dark astronomical site. On the
  other hand, the sky is brighter in the $I$-band, mostly due to the strength
  of the water-vapor Meinel bands.

\item The extinction, measured along the last 4 years, shows a
  seasonal dependency with a typical value of $k_V\sim$0.15 mag in
  Winter time and a wide range of values in Summer, most of them
  restricted to $k_V<$0.4 mag. This seasonal pattern, caused by
  Saharan dust, is similar to the one found in La Palma, but with a
  smaller range of values in Summer time. The analysis of the typical
  extinction coefficients at different wavelenghts for each season
  indicates that the rise of the extinction in Summer is due to an
  increase of Aerosols (dust) in this period of the year.  Due to the
  reduced contribution of the Ozone absorption to the extinction, and
  the stability of the contribution of the Rayleigh scattering along
  the year it was possible to derive an aproximate expression for the
  extinction curve parametrized only by the $V$-band extinction.

\item The fraction of astronomical useful nights, when the weather was good
  enough to allow an acurate measurement of the extinction, was $\sim$70\% of
  the nights in the last 4 years. This fraction is similar to the one found in
  La Palma (Benn \& Ellison 1998). The fraction of these nights that were
  photometric was a $\sim$30\%.

\item The median seeing along the last 2 years was 0.90$\arcsec$, being
  slightly better in Summer (0.87$\arcsec$) than in Winter
  (0.96$\arcsec$). The seeing was better than 1$\arcsec$ in a $\sim$70\% of
  the nights. Although this seeing is slightly worse than in some astronomical
  sites (eg. Mauna Kea, La Palma), it is better than the currently seeing at
  Paranal or MtGraham, two astronomical sites where 10m-like telescope are
  currently in operation or under construction.

\end{itemize}

We conclude that Calar Alto remains a good astronomical site, similar
in many aspects to places where there are 10m-like telescopes under
operation or construction. It will strongly benefit from a sky
protection law that would reduce the light pollution, particularly due
to Mercury and high-pressure Sodium street lamps. Such a law has been
under discussion by the local Andalusian goverment during the last few
years and we hope it will be soon operative.

The fact that Calar Alto is placed in continental Europe is a major
advantage in comparison with other European observatories away from
the continent, since both the operational and development costs are
significantly smaller.

For both reasons we consider that this observatory is a good candidate
for the location of future large aperture optical telescopes.

\section{Acknowledgments}

  SFS thanks the Spanish Plan Nacional de Astronom\'\i a program
  AYA2005-09413-C02-02, of the Spanish Ministery of Education and Science and
  the Plan Andaluz de Investigaci\'on of Junta de Andaluc\'{\i}a as research
  group FQM322.

\newpage

\begin{table}
\begin{center}
\caption{Log of the data-set per night}
\label{tab_data}
\begin{tabular}{llr}\hline
\tableline\tableline
Date & Instrument & wavelegth range\\
\tableline
25/05/05&PPAK&3700-7000\AA\\
26/05/05&PPAK&3700-7000\AA\\
02/06/05&PPAK&3700-6800\AA\\
30/06/05&PPAK&4700-7940\AA\\
03/07/05&PPAK&3700-7100\AA\\
05/07/05&PPAK&3700-7100\AA\\
23/08/06&PPAK&3700-7000\AA\\
\tableline\tableline
Date & Instrument & Filter Bands\\
\tableline
28/08/05&CAFOS&U,B,V,R \& I-bands\\
28/03/06&CAFOS&B,V,R \& I-bands\\
\tableline
\end{tabular}
\end{center}
\end{table}

\begin{table}
\begin{center}
\caption{Properties of the detected emission lines}
\label{tab_line}
\begin{tabular}{llrr}\hline
\tableline\tableline
Line Id & Wavelength & Flux$^*$ & Flux$^{**}$\\
        & (\AA)      &   & R \\
\tableline
HgI     &4047    & 3.5 &  9.6\\
HgI     &4078    & 0.7 &  2.0\\
NaI     &4165,8  & 1.2 &  3.6\\
HgI     &4358    & 6.7 & 22.9\\
NaI     &4420,3  & 0.3 &  1.1\\
TiI     &4511.6  & 3.6 & 13.7\\
NaI     &4665,9  & 1.4 &  5.9\\
HgI     &4827,32 & 0.5 &  2.3\\
NaI     &4978,83 & 1.5 &  7.7\\
NaI     &5149,53 & 0.5 &  2.8\\
NI      &5199    & 0.7 &  4.1\\
ScI     &5351.1  & 3.0 & 19.0\\
HgI     &5461    & 6.4 & 43.1\\
OI      &5577    &32.7 &234.5\\
NaI     &5683,88 & 2.9 & 22.0\\
HgI     &5770,91 & 5.0 & 39.7\\
NaI     &5893,6  & 4.0 & 33.8\\
Broad NaD&5893   &14.9 &126.0\\
NaI     &6154,61 & 0.9 &  8.7\\
OI      &6300    &18.5 &191.2\\
OI      &6364    & 6.5 & 69.3\\
Li      &6708    & 0.3 &  3.7\\
\tableline
\end{tabular}

$*$ in units of 10$^{-16}$ erg~s$^{-1}$~cm$^{-2}$\\

$**$ in units of rayleighs.

\end{center}
\end{table}

\begin{table}
\begin{center}
\caption{Contribution of the light pollution lines to the sky-brightness}
\label{tab_cont}
\begin{tabular}{ll}\hline
\tableline\tableline
Band  & $\Delta$mag\\
$B$   & 0.09 \\
$V$   & 0.16 \\
$R$   & 0.10 \\
3970/331 & 0.08\\
4280/331 & 0.16\\
5510/331 & 0.10\\
5820/331 & 0.33\\
\tableline
\tableline
\end{tabular}
\end{center}
\end{table}

\begin{table}
\begin{center}
\caption{Summary of the night-sky surface brightness}
\label{tab_mag}
\begin{tabular}{lllllll}\hline
\tableline\tableline
Site \& Date &        U   &    B       &    V     &    R    &   I   & Reference\\  
\tableline
02/06/2005$^*$ &  ----       & 22.42$\pm$0.04 & 21.66$\pm$0.15 & 21.04$\pm$0.13 &  ----- & \\
28/08/2005 & 21.81$\pm$0.10 & 22.22$\pm$0.06 & 21.27$\pm$0.18 & 20.46$\pm$0.07 & 18.27$\pm$0.06 & \\
28/03/2006 &  ----       & 22.59$\pm$0.17 & 21.67$\pm$0.13 & 21.04$\pm$0.14 & 19.12$\pm$0.08 & \\
\tableline
 mean       &  21.81$\pm$0.10 &  22.41$\pm$0.15 &  21.53$\pm$0.18 & 
  20.84$\pm$0.27 &  18.70$\pm$0.85  & \\
 air-mass corrected &  22.39$\pm$0.10 &  22.86$\pm$0.03 & 
  22.01$\pm$0.07 &   21.36$\pm$0.11 &  19.25$\pm$0.14 & \\
\tableline\tableline
La Silla      1978 &          &    22.8     &    21.7     &  20.8       & 19.5  & Mattila et al. (1996)\\
Kitt Peak     1987 &          &    22.9     &    21.9     &             &       & Pilachowski et al. (1989)\\
Cerro Tololo  1987 &   22.0   &    22.7     &    21.8     &  20.9       & 19.9  & Walker (1987a,88a)\\
Calar Alto 1990    &   22.2   &    22.6     &    21.5     &  20.6       & 18.7  & Leinert et al. (1995)\\   
La Palma 1990-92   &          &    22.5     &    21.5     &             &       & Benn \& Ellison (1998a,b)\\
La Palma 1994-96   &   22.0   &    22.7     &    21.9     &  21.0       & 20.0  & Benn \& Ellison (1998a,b)\\
Mauna Kea 1995-06  &          &    22.8     &    21.9     &             &       & Krisciunas (1997)\\
Paranal  2000-01   &   22.3   &    22.6     &    21.6     &  20.9       & 19.7  & Patat et al. (2003)\\
MtGraham 2000-01   &   22.38  &    22.86    &    21.72    &  21.19      &       & Taylor et al. (2004)\\ 
Cerro Pachon 2005  &   22.1   &    22.43    &    21.63    &             & 20.3  & Walker \& Schwarz (2007)\\
\tableline
\end{tabular}

$*$ Based on spectrophotometric data obtained with PPAK.

\end{center}
\end{table}

\begin{table}
\begin{center}
\caption{Extinction Coefficients at Calar Alto}
\label{tab_ext}
\begin{tabular}{cll}\hline
\tableline\tableline
Wavelength  & $\kappa_\lambda$ & $\kappa_\lambda$ \\
 (nm)       & Summer           &  Winter \\
\tableline
 380 & 0.505$\pm$0.126 & 0.375$\pm$0.034\\
 436 & 0.324$\pm$0.057 & 0.223$\pm$0.022\\
 500 & 0.216$\pm$0.049 & 0.144$\pm$0.010\\
 671 & 0.109$\pm$0.034 & 0.055$\pm$0.005\\
 880 & 0.065$\pm$0.017 & 0.024$\pm$0.001\\
1020 & 0.049$\pm$0.019 & 0.014$\pm$0.002\\
\tableline
\tableline
\end{tabular}
\end{center}
\end{table}

\begin{table}
\begin{center}
\caption{Contribution of the Rayleigh scattering ($f_1$), Aerosol
  extinction ($f_2$) and Ozone absorption ($f_2$) to the total extinction.}
\label{tab_ecurve}
\begin{tabular}{llllll}\hline
\tableline\tableline
Site & height[m] & $f_1$ & $f_2$ & $f_3$ & reference\\
\tableline
Calar Alto (Winter) & 2168 & 1.02 & 0.94 & 0.29  & this work\\
Calar Alto (Summer) & 2168 & 1.18 & 4.52 & 0.19  &  ``   `` \\
\tableline
Calar Alto (Winter) & 2168 & 1.25 & 1.00 & 0.00  & Hopp \& Fernandez (2002)\\
Lick                & 1290 & 1.25 & 0.80 & 0.00  &   ``            ``   $*$ \\
Palomar             & 1706 & 1.65 & 0.35 & 0.03  &   ``            ``   $*$ \\
KPNO                & 2120 & 1.40 & 0.75 & 0.00  &   ``            ``   $*$ \\
CTIO                & 2215 & 1.50 & 1.10 & 0.00  &   ``            ``   $*$ \\
\tableline
\tableline
\end{tabular}

(*) Data collected by Hopp \& Fernandez (2002) from different studies.
\end{center}
\end{table}

\begin{table}
\begin{center}
\caption{Median seeing compared with other astronomical sites}
\label{tab_seeing}
\begin{tabular}{lrl}\hline
\tableline\tableline
Site & Median seeing & Reference \\
\tableline
Calar Alto (all)    & 0.90$\arcsec$\ \ \  & this work\\
Calar Alto (Winter) & 0.96$\arcsec$\ \ \  & ``    `` \\
Calar Alto (Summer) & 0.87$\arcsec$\ \ \  & ``    `` \\
\tableline
Mauna Kea (1987)    & 0.50$\arcsec$\ \ \  & Racine (1989)\\
La Palma (1997)      & 0.76$\arcsec$\ \ \   & Mu\~noz-Tu\~non et al. (1997)\\
La Silla  (1999)     & 0.79$\arcsec$\ \ \  & ESO webpage$^{*}$\\
Paranal (2005)       & 0.80$\arcsec$\ \ \  & ESO webpage$^{**}$\\
MtGraham (1999-2002) & $\sim$0.97$\arcsec$\ \ \  & Taylor et al. (2004)\\
Paranal (2006)       & $\sim$1.00$\arcsec$\ \ \  & ESO webpage$^{***}$\\
KPNO    (1999)       & $\sim$1.00$\arcsec$\ \ \ & Massey et al. (2000)\\
Lick (1990-1998)     & $\sim$1.90$\arcsec$\ \ \  & MtHamilton webpage$^{****}$\\
\tableline
\tableline
\multicolumn{3}{l}{}\\
\multicolumn{3}{l}{(*) http://www.ls.eso.org/lasilla/seeing/ }\\
\multicolumn{3}{l}{(**) http://www.eso.org/gen-fac/pubs/astclim/paranal/seeing/adaptive-optics/statfwhm.html}\\
\multicolumn{3}{l}{(***) http://www.eso.org/gen-fac/pubs/astclim/paranal/seeing/singstory.html}\\
\multicolumn{3}{l}{(****) https://mthamilton.ucolick.org/techdocs/MH\_weather/obstats/seeing.html}\\
\end{tabular}
\end{center}
\end{table}

\newpage

 \label{lastpage}


\end{document}